\documentclass{amsart}
\usepackage{graphicx}
\theoremstyle{definition}

\theoremstyle{remark}

\numberwithin{equation}{section}

\newcommand{\rc}{\mathbb{R}}

\begin{document}

\title{Entanglement Exchange and Bohmian Mechanics}%
\author[Huggett and Vistarini]{Nick Huggett\\Tiziana Vistarini\\Philosophy, University of Illinois at Chicago}%
\email{huggett@uic.edu, tvista2@uic.edu}%
\thanks{Thanks to the audience at Jon Jarrett's Spring 2009 seminar -- especially Jon.}%
\maketitle %



\begin{quotation}
When two systems, of which we know the states by their respective representatives, enter 
into temporary physical interaction due to known forces between them, and when after a 
time of mutual influence the systems separate again, then they can no longer be described 
in the same way as before, viz. by endowing each of them with a representative of its own. \dots \ 
 \emph{By the interaction} the two representatives have become entangled. Schr\"odinger (1935, 555; our emphasis)
\end{quotation}

\section{Entanglement exchange}
\label{EE}

One of the surprises that the study of quantum entanglement has revealed is that it is not necessary for systems to interact causally in order to evolve from an unentangled state into an entangled one. As the phenomenon of
entanglement exchange (Bennett et al, 1993; Zukowski et al, 1993) shows, it is not required at all; and nor is the effect purely theoretical, since it has
been realised in the laboratory (Zeilinger at al, 1998). The purpose of this paper is to analyze the process within the
framework of Bohmian mechanics, to see how the non-causal `interaction' involved is modeled by its unique account of
the `connectedness' of quantum particles. We hope in this way to illuminate the way in which the `pilot wave' affects particles.

However, the first job is to review entanglement exchange within the `standard' quantum framework, to explain the
phenomenon for those who are unfamiliar with it, and to begin to introduce the notation that we will use in this paper.

Consider a system of four identical particles, each of which has a 2-dimensional `internal' state space $\mathcal{H}_1$, with the
orthonormal basis $ \{|a\rangle, |b\rangle\}$; their joint state space is the 16-dimensional tensor product $\mathcal{H}_4 = \mathcal{H}_1\bigotimes\mathcal{H}_1 \bigotimes\mathcal{H}_1\bigotimes\mathcal{H}_1$. In order to keep our expressions
manageable and readable, we will not use the full bra-ket notation, instead we just use the state labels to indicate
states; thus $a$ will represent the state $|a\rangle$. Additionally, and for the same reason, we will not write out full
tensor products; to keep track of which single particle state belongs in which `slot' of the tensor product, we will use
subscripts. Thus, for example, the state $|a\rangle\otimes|b\rangle\otimes|b\rangle\otimes|a\rangle$ may be written
$a_1b_2b_3a_4$, or $a_1a_4b_2b_3$, or any other permutation.

Now, consider $\mathcal{H}_2 = \mathcal{H}_1\bigotimes\mathcal{H}_1$. The most familiar bases for this space are (in the notation we have adopted) probably $\{a_1a_2, a_1b_2, b_1a_2, b_1b_2\}$ and $\{a_1a_2, a_1b_2+b_1a_2, a_1b_2-b_1a_2, b_1b_2\}$ but, as can be easily checked, it is also spanned by

\begin{equation}
\label{ } \{a_1b_2+b_1a_2, a_1b_2-b_1a_2, a_1a_2+b_1b_2, a_1a_2-b_1b_2\},
\end{equation}

\noindent the unnormalized `Bell basis'. For slots $i$ and $j$ ($i\neq j$) of $ \mathcal{H}_4$, we define normalized Bell states\footnote{Since we are considering only the internal degrees of freedom here, the `Symmetrization Postulate' does not rule out identical particles possessing all these states. It only constrains the \emph{total} state, which involves all degrees of freedom.}:

\begin{eqnarray}
\label{Bexpand}
     \alpha_{i,j} \equiv 1/\sqrt{2}(a_{i} b_{j} + b_{i}a_{j})   &   \gamma_
{i,j} \equiv 1/\sqrt{2}(a_{i}a_{j}+ b_{i}b_{j}) \nonumber\\
      \beta_{i,j} \equiv 1/\sqrt{2}(a_{i} b_{j} - b_{i}a_{j})  &    \delta_
{i,j} \equiv 1/\sqrt{2}(a_{i}a_{j}- b_{i}b_{j}).
\end{eqnarray}

With the basic notation in hand, we can quickly describe what it means to exchange entanglement. Suppose that the system
is in a state in which particles 1 and 2, and 3 and 4 -- \emph{but no others} -- are entangled; specifically let the
initial state be

\begin{equation}
\label{initial}
\Psi = 1/2 (a_{1} b_{2} + b_{1}a_{2}) (a_{3} b_{4} + b_{3}a_{4})= \alpha_{1,2}\alpha_{3,4}.
\end{equation}

$\Psi$ can be decomposed using the Bell basis:

\begin{eqnarray}
\label{initialfac}
\Psi & = & 1/4\cdot\big\{(a_1b_3+b_1a_3)(a_2b_4+b_2a_4)\nonumber\\
&& {} - (a_1b_3-b_1a_3) (a_2b_4-b_2a_4)\nonumber\\
&& {} +(a_1a_3+b_1b_3)(a_2a_4+b_2b_4)\nonumber\\
&& {} - (a_1a_3-b_1b_3) (a_2a_4-b_2b_4)\big\}\nonumber\\
& = & 1/2 \cdot \{\alpha_{1,3}\alpha_{2,4} - \beta_{1,3}\beta_{2,4} + \gamma_{1,3}\gamma_{2,4} - \delta_{1,3}\delta_{2,4}\}.
\end{eqnarray}

We emphasize that in this state neither 1 nor 2 is entangled with 3 or 4, as can clearly be seen by the factorization of $\Psi$  in  (\ref{initial}). But now suppose that a measurement of the Bell
state of 1 and 3 is carried out (i.e., a measurement to determine whether the Bell state of 1 and 3 is $\alpha_{1,3}$,
$\beta_{1,3}$, $\gamma_{1,3}$ or $ \delta_{1,3}$). As a result $\Psi$ will be projected onto one of the four terms of
(\ref{initialfac}), corresponding to the outcome: $ \Psi \to \{\alpha_{1,3}\alpha_{2,4}, \beta_{1,3}\beta_{2,4},
\gamma_ {1,3}\gamma_{2,4}, \delta_{1,3}\delta_{2,4}\}$. But whatever the outcome, particles 2 and 4 will end up in a
Bell state, \emph{and hence be entangled}. The measurement of 1 and 3 alone induces the entanglement of 2 and 4; no interaction, either unitary or involving their measurement, occurs between 2 and 4.

In their paper on the experimental realization of entanglement exchange, Zeilinger et al (1998) remark that entanglement exchange `cause[s] no
conceptual problems [for the Copenhagen interpretation of quantum mechanics] if one accepts that information about
quantum systems is a more basic feature than any possible ``real" properties these systems might have.' The cynic about this interpretation might
respond that if nothing is `really' happening then very little is likely to cause conceptual problems -- except the
underlying problem of what the physical world is like. But of course, that is \emph{the} problem of physics. One may not
like them, but one should surely understand why alternative interpretations -- such as Bohm's -- that tell a story about
`real' properties need to be pursued.

\section{Bohmian Mechanics}

In this section we'll introduce the basic principles of Bohmian mechanics (BM), in part for those unfamiliar with the theory, but also to develop the notation and tools needed for the following section. In so doing we will closely follow the approach developed by Albert (1992) and Barrett (e.g., 2000), to the formulation of the theory by D\"urr, Goldstein and Zanghi, (1992) (see also Goldstein, 2006). Let's start by
analyzing what it means in this account to describe a physical system's state at a certain time and its temporal evolution, by comparison with familiar theories.

In classical mechanics it can be done by describing the positions of the particles forming the system and the momenta they carry. For example, the state at a time of one particle in ordinary space, is described by a point $(q, p)$ in a six-dimensional phase space homeomorphic to $\rc^{6}$, so $q$ in fact represents a point $(x,y,z)$ in 3-space, and $p$ a momentum vector. (In general, for notational ease, we will only explicitly indicate vectors when it is important to recognize their vectorial nature.) The future evolution of that state  in this vector space will be guided by Newton's equations.

In quantum mechanics (QM), the state space is a Hilbert space, and the wave function $\psi$ representing the state of the system is a ray in this space.  This function encodes all the information  about the particle's state and its evolution through time. More precisely, the wave function $\psi(x,t)$ is solution of the Schr\"odinger equation

\begin{equation}
\label{Schrodinger1}
   -\frac{\hbar^{2}}{2m}\nabla^{2}\psi + V\psi = i \hbar \frac
{\partial \psi}{\partial t}.
\end{equation}

While the wavefunction is a function of position, in no obvious way does it specify a point in physical space as the position of the particle; moreover, there is no dependence on such a position in  (\ref{Schrodinger1}). Indeed, the position variable in $\psi$ does not, according to the standard interpretation, refer to a definite position at all. Instead all the theory delivers is the probability of finding the particle in some definite position at a time, provided we perform a measurement at that time: the Born Rule tells us that the probability (density) for a particle's position at any time $t$ is given by $|\psi(x,t)|^{2}$. So, unlike the $q$ of classical mechanics, $x$ does not refer to a particle's actual position at a time, but merely ranges over its possible positions, in a probability function (or better, probability amplitude).

However, if it is true that when we measure a particle's position we get a definite outcome, why can't we just assume that the particle would have been there anyway, even if we did not perform any measurement? The problem of course is the characteristic quantum phenomenon of interference, which involves an extended wavefunction, not one localized at the observed position. So where was the particle before we performed the measurements?  The conventional interpretation of QM would reply: the particle wasn't really in any place. It ended up in  a definite position only because our measurement compelled the particle to have a position.

The crucial feature of the Bohm theory is that it assigns definite particle positions at all times.
We want emphasize this point: a particle's position, $q(t)$, assumes a definite value at any time, because particles have definite positions in this account, whether or not we measure them. For this reason it is often misleadingly thought of as just a revision of classical mechanics. But let's say something more about that later, after we give a more complete description of the theory.

From the wavefunction we get the usual statistical predictions of QM. The $\textit{Distribution Postulate}$, one of the main principles of BM, states that the probability that a particle occupies the position $x$ at the time $t=t_{0}$ is equal to $|\psi(x,t_{0})|^{2} $. To recover the Born Rule we need to ensure that the probability remains $|\psi(x,t)|^{2}$ at times $t\neq t_0$. Imagine that we have an ensemble of $N$ particles all with wavefunction $\psi$, and the same single-particle Schr\"odinger equation, initially distributed so that $\rho(x, t_0) = N\cdot|\psi(x,t_{0})|^{2}$ gives the number (density) of particles at $x$. We want them to move so that $\rho(x, t) = N\cdot|\psi(x,t)|^{2}$ gives the density at any time, in agreement with the Born Rule. Now, a standard textbook result is that the quantum probability current is given by

\begin{equation}
 \vec{j}(x) = \frac{\hbar}{m} \textit{Im}(\psi^{\star}(x)\vec{\nabla} \psi|_{x})
\end{equation}

\noindent (Merzbacher, 1970, pp. 35-9). So if the particle density remains  $N\cdot|\psi(x,t)|^{2}$ as the wavefunction evolves, then $N\cdot \vec{j}$ must be the particle current. But a current is just $\rho\cdot \vec{v}(x)$, where $\vec{v}(x)$ is the velocity of the system (at $x$). In our ensemble, $\rho(x) = N\cdot|\psi(x)|^{2}$, so the Born Rule will be satisfied if, when a particle in the ensemble has position $q$, its velocity is given by

\begin{equation}
\label{v-field1}
\vec{V}[\psi(x)](q) = \frac{\vec{j}(q)}{\rho(q)} =\frac{\hbar}{m}\frac{\textit{Im}(\psi^{\star}(q)\vec{\nabla}\psi|_{q})}{|\psi(q)|^{2}}.\footnote{Note that $\vec{V}$ is a `functional' of the wavefunction and a function of the particle position: that is, a map from a function on space, $\psi(x)$, and position, $q$, to a velocity. The notation of  (\ref{v-field1}) indicates this formal distinction by using different brackets.

That every particle has a well-defined position and velocity does not imply that the Hermitian operators representing position and velocity (or position and momentum) commute in BM. Such operators do not correspond directly to the definite quantities of BM, such as the particle's $q$ and $p$. See Daumer et al (1997) for an account of quantum observables from a Bohmian perspective.}
\end{equation}

$\vec{V}[\psi(x)](q)$ is then the `velocity field' for a single particle system, its postulation by BM as the dynamics for the particle position entails (in conjunction with the Schr\"odinger equation and Distribution Postulate) the Born Rule, and hence all the predictions of QM. (Note that although we derived  (\ref{v-field1}) by considering $N$ particles, it is the equation for a \emph{single} particle; the $N$ particles are merely a \emph{statistical} ensemble, used to treat the probabilities of a single particle being located at this or that place.)

The central role played in Bohm's account  by the notion of particle's position provides us with a concept of system's state that is different in many aspects from the standard interpretation of QM. Although the two accounts have the same statistical predictions, probabilities are used in a different ways. In Bohm's perspective probability is an epistemic tool. The particles always have  definite positions -- from which they evolve in an entirely deterministic way -- and the use of probability is a measure of our ignorance of how they really behave. Of course the probabilities are objective, in the sense that it follows from the distribution postulate that ensembles of identically prepared particles will be statistically distributed according to $|\psi(x)|^2$.

Moreover, the wavefunction $\psi$ contributes to the dynamics of the system in configuration space (in contrast to the standard interpretation of QM in which there are no such degrees of freedom). The equation for the velocity field (\ref{v-field1}) clearly illustrates this point. It determines the velocity of a particle located at $q$ at any time $t$, and so the position at any other times. But $\vec{V}(q)$ is given in terms of $\psi$ and its derivative, and so ultimately it is the wave function that tells us  how a particle evolves to another definite position. (In fact the velocity field is a \emph{local} function of $\psi$ and its derivative, i.e. $\vec{v}(q)$ depends only on the value of $\psi(q)$ and $\vec{\nabla}\psi|_{q}$, a point to which we will return later in this section.) Therefore, the particles are in some sense ``pushed" around  by the wave function on the configuration space -- hence BM is sometimes referred to as the `pilot wave' theory, using de Broglie's expression from the 1927 Solvay Conference (see Goldstein 2006).   It is worth emphasising that the onlyway BM offers the experimentalist to interact with particles is through the wavefunction and velocity field: by a choice of
Hamiltonian or (equivalently) unitary dynamics, \emph{not} by any direct forces or other influences on the particle.

At this point something more should be said about the relation of Bohm's Theory to classical and quantum mechanics. Our position is that BM, rather than a reformulation of classical concepts, should be seen as a particular interpretation of QM, even though the state of a particle is given not just by specifying  its wavefunction, but also by describing the particle position, $q(t)$. First, the particle's  wavefunction evolves according to the standard Schr\"{o}dinger equation (\ref{Schrodinger1}), and particles' definite positions neither appear in $\psi$ nor feature in the Schr\"odinger equation, so the position variable of the wavefunction has much the same significance as before -- a variable in a probability function. Second, since $\psi$ and $q$ fix the velocity of the particle, the linear momentum $p$ is \emph{not} an independent degree of freedom, unlike any of the approaches to classical mechanics. It is true that Bohm's original formulation dresses BM in the guise of a Hamiltonian theory, with a `quantum potential', but that only serves to obscure the crucial innovations of BM. For these reasons, it seems at least as accurate to see BM as an alternative interpretation of QM: it adds structure to the bare bones of QM rather than to classical mechanics. We emphasize this point to stress that BM is not a `reactionary' stance, somehow throwing away crucial insights of QM; how could it if it incorporates QM (as opposed to a philosophically loaded interpretation of QM)?

Now, entanglement exchange, as presented in \S\ref{EE}, involves `internal' degrees of freedom, such as spin, so we need to explain how they are treated in BM. (As a matter of fact, we see no reason that entanglement in spatial degrees of freedom should not be exchanged, but we will stick to the more familiar situation.) Such degrees of freedom are treated exactly as in familiar approaches to QM. That is, the state of a particle is taken to be the (tensor) product of an appropriate internal state and its wavefunction, and the product state evolves according to the usual unitary dynamics.

The only novel feature involving internal degrees of freedom is the dependence of the velocity field on the internal degrees of freedom. The modification to  (\ref{v-field1}) is the simplest possible: the wavefunction is replaced by the product of the internal state and wavefunction, and complex conjugation of the internal state is interpreted as Hermitian conjugation. Thus for an orthonormal set of internal states $\chi_i$ ($i=1,\dots, n$) we have $\chi^*_i\cdot\chi_j = \delta_{i,j}$.\footnote{In this one case $\delta$ is the Kronecker delta, not that defined in  (\ref{Bexpand}).}

Let's work through a simple example to illustrate the function of internal degrees of freedom in BM. So suppose that a system has a Hamiltonian that produces the following evolutions:

\begin{eqnarray}
\label{SG}
\chi_1\psi_0(x) \to \chi_1\psi_1(x) & \mathrm{\ and\ } & \chi_2\psi_0(x) \to \chi_2\psi_2(x)\end{eqnarray}

\noindent where $\psi_1$ and $\psi_2$ have disjoint supports\footnote{$Supp[\psi_{1}] \bigcap   Supp[\psi_{2}]= \emptyset$. $Supp[f] \equiv \{x: f(x) \neq 0\}$, i.e.,
the region in which the function is non-zero. We will make such stipulations about wavefunctions throughout for simplicity, but it is an idealization that makes no essential difference to our results.}. For example, such a system is a simple model of a Stern-Gerlach apparatus: $\chi_1$ and $\chi_2$ represent spin up and down, $\psi_0$ represents the initial position (as the particle enters the apparatus, say), and
 $\psi_1$ and $\psi_2$ represent final locations to the North and South of the apparatus' poles. Such an interaction amounts to a measurement of the internal state, with the result recorded in the position of the particle (which could itself be measured after the experiment). The Born Rule yields zero chance of finding a particle in the region where the wavefunction vanishes, so if the particle starts with internal state $\chi_1$ it will with certainty end up in the support of $\psi_1$, and hence not in the support of $\psi_2$, because they are disjoint (similarly if the initial state is $\chi_2\psi_2$).

The linearity of the quantum dynamics fixes the evolution for a superposition:

\begin{equation}
\label{spinmeas}
\Psi(x) = 1/\sqrt{2}(\chi_1+\chi_2)\psi_0(x) \to 1/\sqrt{2}\big(\chi_1\psi_1(x) + \chi_2\psi_2(x)\big).
\end{equation}

\noindent Once again, the probability for the particle to end up in a region in which the wavefunction vanishes is zero, so it must be in the support of either $\psi_1$ or $\psi_2$ -- by the logic of the previous paragraph, the measurement of the internal degrees of freedom still has an outcome! Now, whether the outcome is 1 or 2 does not depend on the initial internal state, since that is expressed by the LHS of  (\ref{spinmeas}), whatever the result. For the same reason, neither does it depend on the spatial wavefunction. Therefore the result must depend on the initial position, $q(t=0)$, of the particle; but as long as `half' the possible initial positions (i.e., all initial positions in a region over which the integral of $|\Psi(x)|^2$ is $1/2$) end up in the support of $\psi_1$ ($\psi_2$) then the theory will be in agreement with the statistical predictions of QM. (And we re-emphasize that the initial position plus time-dependent wavefunction suffice to determine the motion of the particle.) We won't prove that this is the case, or extend the point to correlations between measurements, but refer the reader to the references  of this section to demonstrate that BM reproduces the statistical predictions of QM regarding internal degrees of freedom as well as position measurements.

Finally, consider the velocity field of the particle in the final state; according to  (\ref{v-field1}) and its modification for internal degrees of freedom discussed above

\begin{eqnarray}
\nonumber \vec{V}[\psi(x)](q) & = & \frac{\hbar}{m}\frac{1/2\cdot\textit{Im}\Big[\big(\chi_1\psi_1(q) + \chi_2\psi_2(q)\big)^*\cdot\vec{\nabla}\big(\chi_1\psi_1 + \chi_2\psi_2\big)|_{q}\Big]}{1/2\cdot|\big(\chi_1\psi_1(q) + \chi_2\psi_2(q)\big)|^{2}}\\
\nonumber & = & \frac{\hbar}{m}\frac{\textit{Im}\Big[\big(\chi^*_1\psi^*_1(q) + \chi^*_2\psi^*_2(q)\big)\cdot\big(\chi_1\vec{\nabla}\psi_1|_{q} + \chi_2\vec{\nabla}\psi_2|_{q}\big)\Big]}{(\chi^*_1\psi^*_1(q) + \chi^*_2\psi^*_2(q)\big)(\chi_1\psi_1(q) + \chi_2\psi_2(q)\big)}\\
& = & \frac{\hbar}{m}\frac{\textit{Im}\Big[\psi^*_1(q) \vec{\nabla}\psi_1|_{q} + \psi^*_2(q)\vec{\nabla}\psi_2|_{q}\Big]}{|\psi_1(q)|^2 + |\psi_2(q)|^2},
\end{eqnarray}

\noindent where the orthonormality of the internal states, $\chi_i$, is used in the final step (e.g., $\chi^*_1\chi_2=0$ and $\chi^*_1\chi_1=1$). Notice that since $\psi_1$ and $\psi_2$ are disjoint $q$ must lie in the support of just one; if, say, $q\in Supp[\psi_1]$ then $\psi_2(q)=0$ and

\begin{equation}
\label{ }
\vec{V}[\psi(x)](q) = \frac{\hbar}{m}\frac{\textit{Im}\Big[\psi^*_1(q) \vec{\nabla}\psi_1|_{q}\Big]}{|\psi_1(q)|^2},
\end{equation}

\noindent (and analogously if $q\in Supp[\psi_2]$ instead). That is, the particle will have the velocity field of a particle with the wavefunction $\psi_1$ (see \ref{v-field1}).

Next, the Bohmian algorithm can be generalized to the case of an $N$ particle system in three dimensions. First, the dynamics for the wavefunction is the standard $N$-particle Schr\"odinger equation, and second, the velocity field becomes

\begin{equation}
\label{v-field}
\vec{V}_Q[\Psi](q_i) = \frac{\hbar}{m_{i}}\frac{\textit{Im}\big(\Psi^{\star}(Q,q_i)\vec{\nabla}_{i}\Psi|_{Q,q_i}\big)}{|\Psi(Q,q_i)|^ {2}}.
\end{equation}
$\vec{V}_Q[\Psi](q_i)$ represents the velocity of the $ith$-particle, located at $q_{i}$, given that the $N$-particle wavefunction (including any internal degrees of freedom) is $\Psi$ and the locations of the other particles are $Q$.\footnote{$\Psi(Q,q_i)$ is of course a suitable permutation of the wavefunction that makes the final argument $q_i$.} This form amounts to an expression for the velocity field for the $i$th particle \emph{given} the positions of the others (and the wavefunction), and we intend the LHS to be read that way. Note that the derivative is only taken with respect to the coordinates of the $i$th particle (hence the subscript on $\vec{\nabla}_i$).

\indent Finally, before we give a Bohmian analysis of entanglement exchange,
it will be helpful to introduce the idea of the \emph{effective
wavefunction}; indeed, this idea will be central to our discussion.
For simplicity, suppose that we have a system of two identical particles (mass $m$)
moving in one dimension; however, it is easy to see that the same
conclusions hold for any number of particles or dimensions. Let the
state of the first particle be $\psi(x)$, and that of the second be $1/\sqrt{2}(
\varphi_{1}(y) + \varphi_{2}(y))$, where $\varphi_{1}$ and $\varphi_{2}
$ have disjoint supports. Thus,

  \begin{equation}
  \Psi(x,y) = 1/\sqrt{2}\cdot\psi(x)(\varphi_{1}(y) + \varphi_{2}(y)).
  \end{equation}

\noindent Of course, under these suppositions the two components of $
\Psi(x,y)$ have disjoint supports in configuration space: there are
no values of $(x,y)$ for which both  $\psi(x)\varphi_{1}(y)$ and  $
\psi(x)\varphi_{2}(y)$ are non-zero.

Now consider the velocity field (according to the 1-dimensional version of  \ref{v-field}) for
the $\psi$-particle, conditional on the $\varphi$-particle being
located at a position $Q$,

  \begin{eqnarray}
V_{Q}[\Psi(x,y)](q) = \frac{\hbar}{m}\frac{\textit{Im}(\Psi^{\star}(q,Q)
\cdot\partial_x\Psi|_{Q,q})}{|\Psi(q,Q)|^{2}}\\
\nonumber \\
= \frac{\hbar}{m}\frac{\textit{Im}\{[\psi(q)(\varphi_{1}(Q) + \varphi_
{2}(Q))]\cdot\partial_x[\psi(q)(\varphi_{1}(Q) + \varphi_{2}(Q))]|_{Q,q}
\}}{|[\psi(q)(\varphi_{1}(Q) + \varphi_{2}(Q))]|^{2}}.\\
\nonumber
  \end{eqnarray}

\noindent By the Born Rule, the particles must have
positions in the support of one of the two components of $\Psi$: the
probability of their being in region where $\Psi$ is zero is of
course $|\Psi|^2=0$. Suppose that $Q\in Supp[\varphi_1]$, hence by the disjointness of the packets, $Q\notin Supp[\varphi_2]$ and $\varphi_2(Q) = 0$. So the expression immediately simplifies to,

   \begin{equation}
V_{Q}[\Psi](q) = \frac{\hbar}{m}\frac{\textit{Im}\{\psi(q)\varphi_{1}
(Q)\cdot\partial_x[\psi(q)(\varphi_{1}(y)]|_{Q,q}\}}{|\psi(q)\varphi_
{1}(Q)|^{2}}.
  \end{equation}

\noindent But inspection of  (\ref{v-field}) reveals that this is
precisely the velocity field that would obtain if the \emph{entire}
wavefunction were $\psi(x)\varphi_{1}(y)$. That is, if $Q \in Supp
[\varphi_{i}]$, $i=1,2$, then

  \begin{equation}
V_{Q}[\Psi](q) = V_{Q}[\psi\varphi_{i}](q).
  \end{equation}

\noindent Effectively, only the packet in whose support the system
lies contributes to the velocity field.

It's easy to see why this result should hold: the velocity field is a
\emph{local} functional of the wavefunction in configuration space,
since its value at $(q,Q)$ depends only on $\Psi$ and its derivative at $(q,Q)$.\footnote{To see what we mean, compare  (\ref{v-field}) with an example of a \emph{non-local} functional such as
$f[\psi(y)](x) = \int_\infty\mathrm{d}y \  \psi(y)\cdot e^{-(x-y)^2}$, which
depends on values of $\psi$ everywhere.} So, \emph{conditional on a
system being located inside the support of a wavepacket}, the
velocity field is independent of any wavepackets with disjoint
supports, because they are necessarily not local to the position of the
system.\footnote{Note that in addition to locality, part of the point
is that if the system is located in some region, then only the
velocity field in that region is relevant to the position dynamics.
We will be concerned with how the particle does move, not how it
would if it were in a different packet.} So, quite generally, if the
wavefunction of a system is the sum of components with disjoint
supports in configuration space, then only the one in whose support
the system is located will contribute to the velocity field; that
component is hence termed the \emph{effective} wavefunction.

The effective wavefunction is central to the Bohmian interpretation of wavefunction `collapse'. Suppose a system decomposes into a measurement apparatus ($A$) and measured subsystem ($S$), and that the different outcomes of a measurement correspond to disjoint wavepackets for the apparatus. For instance, suppose $S$ has possible states $\psi^S|+\rangle$ and $\psi^S|-\rangle$, and measurements of the former leave A in the state $\psi^A_+$, while measurements of the former leave A in $\psi^A_-$, where $\psi^A_+$ and $\psi^A_-$ have disjoint supports -- they correspond to distinct positions for a pointer, say. Then (by linearity) a measurement of $S$ in state $\psi^S(|+\rangle+|-\rangle)$ will yield a state $\psi^A_+\psi^S|+\rangle + \psi^A_-\psi^S|-\rangle$. But the pointer particles can only be in the support of one of the wavepackets. That packet is thus the effective one, and, for the reasons discussed, the system will behave as if that were the entire wavefunction -- as if the wavefunction had collapsed into the wavepacket corresponding to the observed outcome!

It is worth noting (with Albert, 1992, 164-9) that
BM has an implicit supposition to the effect that all that can be
directly observed about the position of a particle is its effective
probability distribution $|\Psi_{eff}|^2$.\footnote
{That the effective wavefunction can be determined is certainly
correct -- suitable measurements will show in which packet a particle
ends up -- but why we should directly observe no more is less clear.
One hypothesis is that the mind's belief states supervene on
effective brain states, not on the exact configuration. We will not
pursue this issue here, as it is not relevant for the treatment of
entanglement exchange.} It follows -- since the effective wavefunction is identical with the collapsed wavefunction of the conventional interpretation of QM -- that that is
all that can be known by any means, implying that the \emph{only}
empirical consequences of the theory are the usual statistical
predictions of QM, not statements about exact positions.\footnote{To be more precise, that agreement is only `for all practical purposes'; since the collapse is only effective, there is always the in-principle possibility of the `empty' parts of the wavefunction causing interference effects, something ruled out by a true collapse. However, decoherence essentially guarantees that such effects will not be seen. See Albert (1992, 161-4).}

\section{Entanglement Exchange in the Bohm Theory}

Not surprisingly, given the discussion of the previous section, the formal treatment of this problem in Bohm's conceptual framework shows different features from those presented in \S \ref{EE}. These formal differences are actually the vehicle of the different physical insights behind the two interpretations of quantum phenomena. First, the definition of a physical system in BM does not include just the particles whose behaviour is under observation, but also the measurement apparatus -- in the case of entanglement exchange, a device that we'll call the ``Bell-ometer". Second, as we explained in the previous section, in BM every particle has a determinate position, whether or not we have measured it. Moreover, the position of a particle has a temporal evolution guided by the `pilot wave' through (\ref{v-field1}). Hence we have to take the spatial wavefunction explicitly into account in the treatment, not just the internal degrees of freedom.

This section will give a full formal treatment of entanglement exchange (in a simple model at least). The calculations are elementary, though the expressions are a little unwieldy (we have five subsystems to consider). In the following section we will analyse the treatment further in a more intuitive, pictorial manner, focussing on the most essential features.

Consider the initial state of the system. In addition to the `internal' part of the state given in (\ref{initial}), we suppose that the four particles ($i=1\mathrm{-}4$) have associated wavefunctions, $\phi_{i}(\vec{x_{i}})$, with mutually disjunct supports. The initial state of the Bell-ometer is in turn represented by a wavefunction $\psi_{0}(x)$ whose argument is a one-dimensional variable representing the possible positions of the pointer. (Of course the pointer is composed of many atoms, but they are in a bound state, so we suppose that they only have one relevant degree of freedom: $x$ represents the position of the centre of mass.) See figure \ref {Bex}. Thus (suppressing unambiguous arguments for simplicity of expression) the total initial state is given by:

\begin{figure}
\begin{center}
\includegraphics[width=4in]{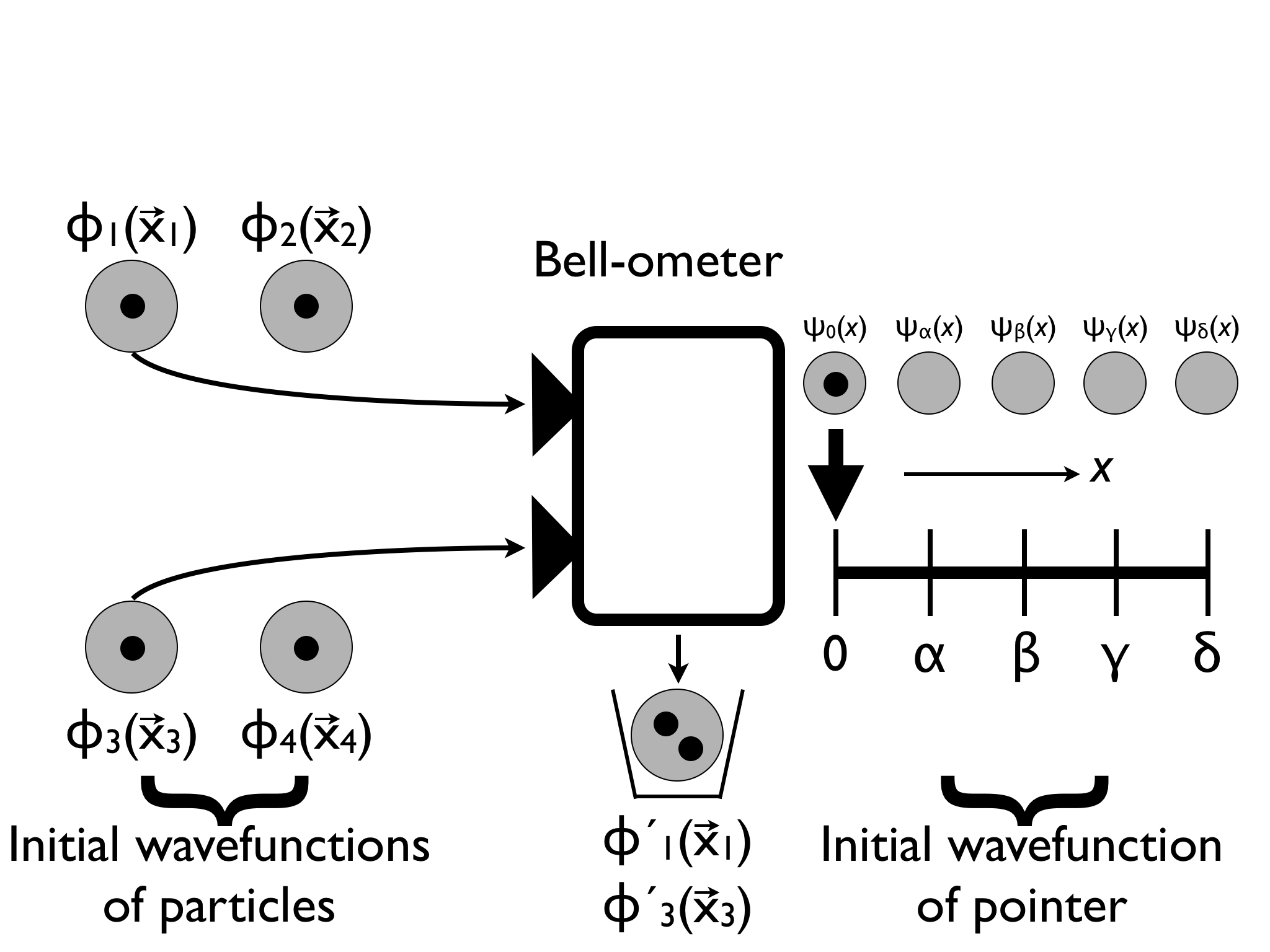}
\caption{ }
\label{Bex}
\end{center}
\end{figure}

\begin{eqnarray}
\label{prebell}
\Psi(t=0) & = & \frac{1}{2}(a_{1} b_{2} + b_{2}a_{1}) (a_{3} b_{4} + b_{3} a_{4})\phi_{1}\phi_{2}\phi_{3}\phi_{4}\psi_{0} \nonumber \\
& = & \frac{1}{2}(\alpha_{1,3} \alpha_{2,4} - \beta_{1,3} \beta_{2,4} + \gamma_{1,3}\gamma_{2,4} -
\delta_{1,3}\delta_{2,4})\phi_{1}\phi_{2} \phi_{3}\phi_{4}\psi_{0},
\end{eqnarray}

\noindent by (\ref{Bexpand}).

Now consider the operation of the Bell-ometer in more detail. In addition to $\psi_{0}$, the `ready state' of the device, there are four other possible wavefunctions for the pointer: $\psi_{\alpha}$ ,$\psi_{\beta}$, $\psi_{\gamma}$ and $\psi_{\delta}$. Each is centered at a different point along a scale, and the location of the pointer at any such point indicates the outcome of the experiment (for convenience, we take all five pointer wavefunctions to have mutually disjoint supports). The supports of the wavefunctions are indicated against their scale in figure \ref{Bex}.

The Bell-ometer then works by feeding in a pair of particles, $i$ and $j$ (particles $1$ and $3$ are shown entering the device). The operation of the joint system of particles and Bell-ometer is then such that -- i.e., has a Hamiltonian such that -- if the internal part of the joint state of the particles is $\alpha_{i,j}$ and the system, including the pointer, is in its ready state, then the unitary Schr\"odinger evolution will lead to a final state for the whole system of $\alpha_{i,j}\varphi^\prime_i\varphi^\prime_j\psi_\alpha$, and so on:

\begin{eqnarray}
\label{bellometer}
\nonumber \alpha_{i,j}\varphi_i\varphi_j\psi_0 \to \alpha_{i,j}\varphi^\prime_i\varphi^\prime_j\psi_\alpha & & \beta_{i,j}\varphi_i\varphi_j\psi_0 \to \beta_{i,j}\varphi^\prime_i\varphi^\prime_j\psi_\beta\\
\gamma_{i,j}\varphi_i\varphi_j\psi_0 \to \gamma_{i,j}\varphi^\prime_i\varphi^\prime_j\psi_\gamma & & \delta_{i,j}\varphi_i\varphi_j\psi_0 \to \delta_{i,j}\varphi^\prime_i\varphi^\prime_j\psi_\delta.
\end{eqnarray}

\noindent $\varphi^\prime_{i/j}(\vec{x})$ can be thought of as wavefunctions located in a particle `dustbin'. They simply indicate that the measured particles are discarded after the experiment and their locations are not relevant for the remainder of our discussion. In reality, the outcome of the experiment may be recorded in the location of the particles as well as the pointer, but in our model the position of the Bell-ometer's pointer represents all the spatial differences that might result from different measurement. Since, by the Born Rule, the particles have zero probability of being outside the support of the wavefunction, they will certainly be located in the dustbin, as the figure shows.

Because the device has the dynamics specified by  (\ref{bellometer}), if the particles are initially in the state $\alpha_{i,j}$ ($\beta_{i,j},\  \dots$) , then \emph{with certainty} the final state for the pointer will be $\varphi_\alpha$ ($\varphi_\beta,\ \dots$). That is, the device indeed measures the Bell state of the particles. This device is clearly theoretically possible, in the sense that there exists a unitary operator with the desired action. But we wish to emphasize again that Zeilinger et al (1998) built and operated a device that effectively operates in just this way in their experiments: the Bell-ometer is a laboratory reality, not just a theoretical possibility.

Finally, since the dynamics is of the standard, unitary, Schr\"{o}dinger kind, it is linear. Thus if the initial state of the particles is a superposition of Bell states, then the outcome will involve a superposition of pointer wavefunctions:

\begin{eqnarray}
\label{bellsup}& & (c_\alpha\alpha_{i,j} + c_\beta\beta_{i,j} + c_\gamma\gamma_{i,j} + c_\delta\delta_{i,j})\varphi_i\varphi_j\psi_0 \quad \to\\ 
\nonumber& & (c_\alpha\alpha_{i,j}\psi_\alpha + c_\beta\beta_{i,j}\psi_\beta + c_\gamma\gamma_{i,j}\psi_\gamma + c_\delta\delta_{i,j}\psi_\delta)\varphi_i^{'}\varphi_j^{'}.
\end{eqnarray}

\noindent However, the pointer can only end up in one position, which must be inside the support of its wavefunction.  The Born Rule gives zero probability for being elsewhere: \emph{it must yield a definite outcome} even in this case. As we discussed in the previous section, that location will pick out a term from the RHS of  (\ref{bellsup}) as the effective wavefunction, and there will thus be an effective collapse.

So back to our analysis of entanglement exchange. Suppose that at $t=1$ particles 1 and 3 are fed into the device and their Bell state measured (as shown in figure \ref{Bex}). Then by our discussion of the Bell-ometer's operation, 

\begin{equation}
\label{postbell}
\Psi(t=1) = \frac{1}{2}(\alpha_{1,3} \alpha_{2,4}\psi_{\alpha}- \beta_{1,3} \beta_{2,4}\psi_{\beta} +
\gamma_{1,3}\gamma_{2,4}\psi_ {\gamma} - \delta_{1,3}\delta_{2,4}\psi_{\delta})\phi_{1}^{'}\phi_{2}
\phi_{3}^{'}\phi_{4}.
\end{equation}

Suppose that the result of the Bell measurement is $\alpha$ (it should be clear that no loss of generality is involved,
because exactly symmetric considerations follow in the case of the other possible outcomes). In this case, $x\in Supp[\psi_\alpha]$ (where $x$ is the definite position of the pointer), and 

\begin{equation}
\label{ } \psi_{\beta}(x)=\psi_{\gamma}(x)=\psi_{\delta}(x)=0
\end{equation}

\noindent because of the mutually disjoint supports of possible pointer wavefunctions. (Note that to avoid a proliferation of variables we are letting $x$, etc, do double duty, both as the wavefunction variable, and as the definite position, which we previously denoted $q$. The context and role in the various expressions disambiguates; we trust that the discussion of the previous section makes the two roles for position clear.) It follows that none of the terms involving those wavefunctions contribute to the velocity field of the system, and so the effective wavefunction is

\begin{equation}
\label{ } \Psi_{eff}(t=1) = \alpha_{1,3} \alpha_{2,4}\psi_{\alpha}\phi_{1}^{'} \phi_{2}\phi_{3}^{'}\phi_{4}.
\end{equation}

Next (at $t=2$) a measurement of the spin of particle 2 is made. The device used is very simple, the Stern-Gerlach apparatus we modeled in the previous section: if a particle with spin a (respectively, b) enters the device, it leaves to the North
(respectively, South) of the device, and ends with wavefunction $\phi^{a}(\vec{x})$ ($\phi^{b}
(\vec{x}))$; once again these wavefunctions have disjoint supports. Thus the spin is
recorded in the position of the particle. At the end of this measurement the effective state will be

\begin{equation}
\Psi_{eff}(t=2) = \alpha_{1,3}(a_{2}b_{4}\phi_{2}^{a} + b_{2}a_{4}
\phi_{2}^{b})\psi_{\alpha}\phi_{1}^{'}\phi_{3}^{'}\phi_{4}.
\end{equation}

Finally, (at $t=3$) the spin of particle 4 is measured, using a second spin measurement device. This procedure exhibits the entanglement of 2 and 4
(produced by the measurement of the Bell state of 1 and 3) because the outcome is perfectly anti-correlated with that
of the spin measurement of 2, as we shall now see.

The post-measurement state will be

\begin{equation}
\Psi_{eff}(t=3 ) = \alpha_{1,3}(a_{2}b_{4}\phi_{2}^{a}\phi_{4}^{b} +
b_{2}a_{4}\phi_{2}^{b}\phi_{4}^{a})\psi_{\alpha}(x)\phi_{1}^{'}\phi_{3}^{'}.
\end{equation}

Now we use the equation for the velocity field (\ref{v-field}) to determine the velocity of particle 4 as a function
of its position, given the outcome of the Bell-ometer experiment and for a specified position of particle 2, $x_2$.

\begin{eqnarray}
\lefteqn{V_{\vec{x}_{2}}[\Psi](\vec{x}_{4}) =} \nonumber  \\
&& \textit{Im}\Big\{\alpha^{\ast}_{13}\Big(a_{2}^{\ast}b_{4}^{\ast}\cdot(\phi_{2}^{a})^{\ast} (\phi_{4}^{b})^{\ast}+ b_{2}^{\ast}a_{4}^{\ast}\cdot(\phi_{2}^ {b})^{\ast}(\phi_{4}^{a})^{\ast}\Big)\psi^{\ast}_{\alpha}\cdot(\phi_{1}^ {'})^{\ast}\cdot(\phi_{3}^{'})^{\ast}\cdot \nonumber \\
&& \hspace{48pt} \frac
{\partial_{x_{4}}\Big[ \alpha_{13}\Big(a_{2}b_{4}\phi_{2}^{a}\phi_{4}^{b}(x_{4})+ b_{2}a_{4}\phi_{2}^{b}\phi_{4}^{a}(x_{4})\Big)\psi_{\alpha}\phi_{1}^ {'}\phi_{3}^{'}\Big] \Big\}}
{\big|\alpha_{13}\Big(a_{2}b_{4}\phi_{2}^{a}\phi_{4}^{b}+ b_{2}a_{4}\phi_{2}^{b}\phi_{4}^{a}\Big)\psi_{\alpha}\phi_{1}^{'}\phi_{3}^{'}\big|^{2}}
\end{eqnarray}

Using the orthonormality of $a$ and $b$ and the fact that only $\phi^{a/b}_4$ are functions of $x_4$ (so, e.g., $\partial_{x_{4}}\phi^{a}_2=0$) the above expression simplifies to,


\begin{eqnarray}
V_{\vec{x}_{2}}[\Psi](\vec{x}_{4}) & = &
\frac
{|\alpha_{13}|^{2}|\phi_{1}^ {'}|^{2}|\phi_{2}^{'}|^{2}|\psi_{\alpha}|^{2}\textit{Im}\big\{|\phi_{2}^{a}|^{2}(\phi_{4}^{b})^{\ast}\partial_{x_{4}} \phi_{4}^{b}+|\phi_{2}^{b}|^{2}(\phi_{4}^{a})^{\ast}\partial_{x_{4}} \phi_{4}^{a}\big\}}
{|\alpha_{13}|^{2}|\phi_{1}^{'}|^{2}|\phi_{2}^{'}|^{2}|\psi_{\alpha}|^{2}(|\phi_{2}^{a}|^ {2}|\phi_{4}^{b}|^{2} +|\phi_{2}^{b}|^{2}|\phi_{4}^{a}|^{2})} \nonumber \\
\nonumber \\
& = & \frac
{\textit{Im}\big\{|\phi_{2}^{a}|^{2}(\phi_{4}^{b})^{\ast}\partial_ {x_{4}}\phi_{4}^{b}+
|\phi_{2}^{b}|^{2}(\phi_{4}^{a})^{\ast}\partial_ {x_{4}}\phi_{4}^{a}\big\}}
{|\phi_{2}^{a}|^{2}|\phi_{4}^{b}|^{2} +|\phi_{2}^{b}|^{2}|\phi_{4}^{a}|^{2}}
\end{eqnarray}
Now suppose that the outcome of the measurement of particle 2's spin was $a$ (again, clearly symmetric results obtain if
we suppose it was $b$). That is to say that the position of 2, $\vec{x}_2$, is outside the support of
$\phi_{2}^{b}$: hence $\phi_{2}^{b}(\vec{x}_ {2})= 0$. In that case we find that

\begin{equation}
V_{\vec{x}_{2}}[\Psi](\vec{x}_{4}) =\frac{\textit{Im}[(\phi_{4}^{b})^
{\ast}\partial_{x_{4}}\phi_{4}^{b}]}{|\phi_{4}^{b}|^{2}}.
\end{equation}

In other words, the velocity field of particle 4 is the same as if its state were

\begin{equation}
\label{ } \psi = b_4\phi^b_4(\vec{x}_4)
\end{equation}

\noindent namely that of a $b$-spin particle -- even though it is in fact in an entangled state. Thus it is
anti-correlated with the spin of 2 --namely $a$ -- as advertised. Moreover, that it behaves as a spin-$b$ particle is determined
completely by the outcome of the two previous measurements.

\section{How do the Particles Get Entangled?}

\begin{figure}
\begin{center}
\includegraphics[width=4in]{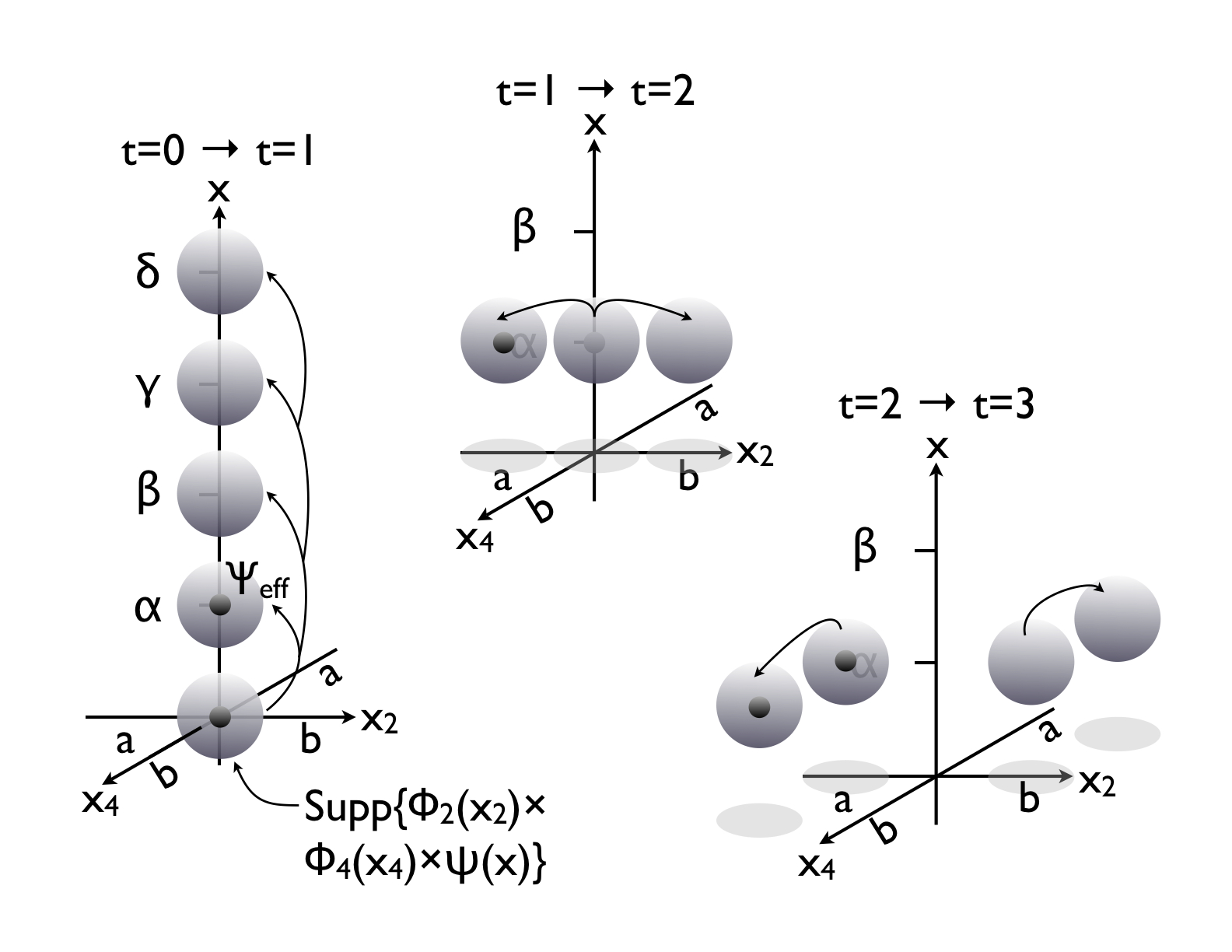}
\caption{ }
\label{EEfig}
\end{center}
\end{figure}

Our system, composed of four particles and a pointer in 3-space, lives in a 15-dimensional configuration space (in addition there are of course the internal degrees of freedom). However, to visualize what happens in the interactions we need only really picture particles 2 and 4 (those that become entangled by the Bell-ometer measurement) and the pointer; in addition, we can simplify by just looking at their movements in one, representative, dimension each. Thus in figure \ref{EEfig}, the vertical axis represents the line along which the pointer moves. The horizontal axis represents the motion of particle 2 parallel to a line through the magnetic poles of the Stern-Gerlach spin measurement device; motion to the left represents a measurement of spin $a$, that to the right of spin $b$. Similarly for the axis perpendicular to the page and particle 4. The three transitions treated, $t=0 \to t=1$ (the Bell-ometer measurement), $t=1 \to t=2$ (the measurement of the spin of 2), and $t=2 \to t=3$ (the measurement of the spin of 4), are all shown. In each case both the evolution of the wavefunction (or effective wavefunction) and the location in configuration space of the system are shown, the former by shading the region(s) in which the wavefunction has support, and the latter by a dot. It will be convenient to think of the whole system as a particle -- the `system particle' -- moving (along with its wavefunction) in $(x,x_2,x_4)$--space.

As in the conventional interpretation of QM, entanglement is exchanged by the Bell state measurement ($t=0 \to t=1$). Afterwards, the system particle is guaranteed (by the Born Rule) to be located in the support of a wavepacket associated with an internal state in which particles 2 and 4 are either perfectly anticorrelated (for outcomes $\alpha$ or $\beta$) or perfectly correlated ($\gamma$ or $\delta$) -- in our example, we considered $\alpha$. But under such circumstances the packet within which the system is located is the effective wavefunction; in our example, a state in which 2 and 4 are in the entangled internal state $\alpha_{2,4}$. As we showed, the system will now behave exactly as if it were in the effective state, and so 2 and 4 will exhibit the appropriate correlations.

Indeed, the second two steps pictured simply show how the correlation is manifested: the measurement of the spin of 2 splits its wavepacket into two, according to the possible outcomes. The subsequent measurement of 4 splits its packet, but the split is anticorrelated with the wavepackets of 2, so that the result has supports only around $(x_2=a, x_4=b)$ and $(x_2=b, x_4=a)$. Of course the system particle can only be located within the support of one disjoint packet, so the result of the measurements is one of the two possible anticorrelated outcomes. However, it is important to stress that these second two steps follow directly from the entanglement between 2 and 4 in the effective state; as such they simply illustrate how any pair of particles in the state $\alpha$ would behave. So they are not a novel feature of entanglement exchange, but just a standard quantum correlation in BM, as discussed for instance in Albert (1992, pp. 155-63).

So let us focus on the first step. As we said, particles 2 and 4 end up entangled because the wavefunction splits into packets, each of which corresponds to an entangled internal state, and because the system must lie within the support of one of the packets -- making it the effective wavefunction. We wish to make a couple of remarks about this process, which we hope shed some light on the Bohmian interpretation of QM.

First, consider the full wavefunction immediately after the Bell measurement (i.e., at $t=1$):  (\ref{postbell}). The internal state of particles 2 and 4 is unentangled; in fact, they are completely uncorrelated. For instance, according to the standard QM algorithm both the unconditional probability that 4 has spin $a$ and the probability it does conditional on 2 having spin $a$ are $1/2$.\footnote{The internal state of 2 and 4 is the reduced density matrix, $\rho_{2,4}$ obtained by tracing over all other degrees of freedom, for all the particles and pointer. If $\rho_\alpha$ is the density matrix for state $\alpha$, etc, then $\rho_{2,4} = 1/4(\rho_\alpha+\rho_\beta+\rho_\gamma+\rho_\delta)$. A simple calculation shows that the RHS is equal to $1/4(\rho_{aa}+\rho_{ab}+\rho_{ba}+\rho_{bb})$, where $\rho_{aa}$ is the density matrix for the state $aa$, etc. Since each of the matrices in this final expression is unentangled, the statistics are those of a mixture of unentangled particles.} So 2 and 4 are not entangled in the full state, only in the effective state. Moreover, the nature of the entanglement -- whether 2 and 4 are correlated or anticorrelated -- also depends on the effective wavefunction. But of course that depends not only on the wavefunction, but on the \emph{position} of the system particle. That is, in BM entanglement is not simply a feature of wavefunctions, but of wavefunctions and positions.

Second, note that no change occurs in either the wavefunctions or internal states of 2 and 4 during the Bell measurement: compare (\ref{prebell}) with (\ref{postbell}), and inspect figure \ref{EEfig}. And yet after the Bell measurement, 2 and 4 are correlated. This feature is a reflection of the non-locality (with respect to physical space) of the theory -- of course, according to standard QM also the 2-4 entanglement can arise from an arbitrarily distant measurement of 1 and 3. More interesting, it demonstrates that the entanglement should be thought of as a property of the whole system, because it arises only through changes in the rest of the system, especially the evolution of the pointer, whose wavefunction and position does change. To make the point even more vivid, suppose that the Bell-ometer measurement was carried out on Mars, and in such a way that the apparatus was screened off from all external degrees of freedom, to prevent any decoherence. Then the correlations of 2 and 4 on Earth could be destroyed by merely recombining the wavepackets of the pointer. Suppose $1/2(\psi_\alpha+\psi_\beta+\psi_\gamma+\psi_\delta) \to \psi_0$, so that

\begin{eqnarray}
\label{ }
\nonumber \Psi(t=1) & \to & \frac{1}{2}(\alpha_{1,3} \alpha_{2,4} - \beta_{1,3} \beta_{2,4} + \gamma_{1,3}\gamma_{2,4} - \delta_{1,3}\delta_{2,4})\phi^\prime_{1}\phi_{2} \phi^\prime_{3}\phi_{4}\psi_{0}\\
& = & \alpha_{1,2}\alpha_{3,4}\phi^\prime_{1}\phi_{2} \phi^\prime_{3}\phi_{4}\psi_{0}
\end{eqnarray}
(using \ref{initialfac}), which is a state in which 2 and 4 are unentangled. That is, the particles are not simply correlated with each other, but with respect to the whole system; the correlations depend on the state of the distant pointer. Similar points hold in other theories in which there is no wavefunction collapse; what is unique to BM is the way that the positions are also relevant.
\\

\noindent\textbf{Bibliography}\\
\nocite{*}

D. Albert, (1992) \textit{Quantum Mechanics and Experience}, Harvard University Press.\\

J. Barrett, (2000) ``The Persistence of Memory: Surreal Trajectories in Bohm Theory", \textit{Philosophy of
Science} 67: 680-703.\\

C. H. Bennett, G. Brassard, C. CrŽpeau, R. Jozsa, A. Peres, and W. K. Wootters (1993) ``Teleporting an Unknown Quantum State via Dual Classical and Einstein-Podolsky-Rosen Channels", \textit{Physical Review Letters} 70: 1895-99.\\

M. Daumer, D. D\"urr, S. Goldstein, and N. Zanghi (1997) ``Naive Realism About Operators", \textit{Erkenntnis} 45: 379-97.\\

D. D\"{u}rr, S. Goldstein, N. Zanghi (1992) ``Quantum Equilibrium and the Origin of Absolute Uncertainty", \textit{Journal of Statistical Physics} 67: 1-75.\\

S. Goldstein (2006) ``Bohmian Mechanics", \textit{The Stanford Encyclopedia of Philosophy} (Spring 2009 Edition), Edward N. Zalta (ed.), URL = $<$http://plato.stanford.edu\\/archives/spr2009/entries/qm-bohm/$>$.\\

E. Merzbacher, (1970) \textit{Quantum Mechanics}, Wiley International Edition.\\

E. Schr\"odinger (1935) ``Discussion of Probability Relations Between Separated Systems", 
\textit{Proceedings of the Cambridge Philosophical Society} 31: 555-63.\\

A. Zeilinger, H. Weinfurter, D. Bouwmeester , J-W. Pan, (1998) ``Experimental Entanglement Swapping: Entangling Photons That Never Interacted", \textit{Physical Review Letters} 80: 3891-94.\\

M. Zukowski, A. Zeilinger, M. A. Horne, and A. Ekert (1993) ``Event-Ready-Detectors'' Bell Experiment via Entanglement Swapping", \textit{Physical Review Letters} 71: 4287-90.\\

\end{document}